\documentclass[aps,prd,onecolumn,amsmath,11pt,superscriptaddress,floatfix,nofootinbib,preprintnumbers]{revtex4-2}

\usepackage{mathrsfs}
\usepackage{amsfonts}
\usepackage{amsmath,amssymb,bm}
\usepackage{array}
\usepackage{verbatim}
\usepackage{epsfig}
\usepackage{graphicx}
\usepackage{hyperref}
\hypersetup{colorlinks, linkcolor = [rgb]{0, 0, 0.5}, citecolor = [rgb]{0,0.0,0.5}, urlcolor = [rgb]{0,0.0,0.5}}
\usepackage[normalem]{ulem}
\usepackage{xcolor}
\usepackage{ulem}
\usepackage{multirow}

\usepackage{comment}
\usepackage{subcaption}
\usepackage{graphicx}
\graphicspath{{./figs/}}
\usepackage{dcolumn}
\usepackage{bm}
\usepackage{slashed}
\usepackage{siunitx}
\usepackage{ulem,xpatch}
\usepackage{hyperref}
\usepackage[mathlines]{lineno}
\usepackage{comment}
\usepackage{soul}
\usepackage{xcolor}
\usepackage{xfrac}

\allowdisplaybreaks[4]

\begin{document}

\title{Studying Bremsstrahlung in Polarized Background Field at EIC and EicC}
\author{Cong Li}
\affiliation{School of Information Engineering, Zhejiang Ocean University, Zhoushan, Zhejiang, China}

%


\begin{abstract}

We present a quantum-field-theoretical study of bremsstrahlung from an electron propagating in a linearly polarized background photon field. 
Starting from the photon two-point correlator, the background-modified photon propagator is parameterized via transverse-momentum–dependent photon distributions. 
Coulomb correction is incorporated through a gauge-link formalism and soft-photon radiation is resummed with a Sudakov factor, yielding an analytic form of the polarized Bethe--Heitler spectrum. 
Numerical illustrations show a characteristic $\cos 2\phi$ azimuthal modulation of a few percent in the differential distribution, with Sudakov suppression at low transverse momentum and mild Coulomb-induced enhancement at larger $q_\perp$.

\end{abstract}

\maketitle
\section{Introduction}
\label{s.intro}
Bremsstrahlung radiation, describing photon emission from an accelerated electron, is one of the most fundamental processes in QED. In its conventional theoretical treatment, the radiation spectrum depends solely on the coupling constant $\alpha_{\mathrm{em}}$ and on the kinematics of the scattering process - especially in particle physics — without carrying any explicit information about the background field itself. In contrast, when the electron moves through a polarized background field, the emission pattern should be influenced by the field’s internal polarization and spatial structure. The present study focuses on this latter situation, in which the information about the background field should be cast into the radiation process.

In optical and laser physics, similar situations have been extensively studied. Despite significant progress, bremsstrahlung radiation in polarized background fields still faces several key limitations and unresolved challenges. One major issue lies in the complexity of the background field itself. Most existing studies assume a uniform and stationary background field, whereas real-world polarized background fields exhibit spatiotemporal variations \cite{PhysRevLett.108.165006,Ritus1985,DiPiazza2012}. Examples include the time-dependent nature of laser pulse fields and the transient electromagnetic fields generated in heavy-ion collisions. Another critical gap is the lack of systematic higher-order corrections. In strong background fields, electrons may undergo multiple scatterings, yet the precise impact of these interactions on bremsstrahlung radiation remains to be fully quantified \cite{Baier_2005,PhysRevA.83.022101}.  In experimental settings, generating and precisely controlling polarized background fields, especially in high-energy particle collision environments, present significant technical challenges \cite{79.1626,Bulanov2015}.

Our research aims to address these three challenges. First, we parameterize the photon propagator in a polarized background field, allowing the background field to be incorporated into our theoretical framework in terms of photon distribution functions. This represents a transition from the classical to the quantum description. Previous approaches have treated the background field as an external electromagnetic one, which becomes significantly more complex when polarization is involved. Our method offers a more systematic way to handle this complexity. Second, we introduce a formalism to account for the multiple scatterings of the final-state electron within the background field. We achieve this by integrating the multiple scatterings into a gauge link, effectively absorbing their effects into the initial-state photon correlation function. To ensure a well-defined and experimentally relevant setting, we employ a well-studied, stable, and well-characterized photon background field as the foundation of our analysis. In our previous research \cite{Li1,Li2}, we investigated the photon distribution outside the nucleus and found that it is linearly polarized. Moreover, this background can be accurately described using the equivalent photon approximation (EPA).  

Based on the above, we analyze the Bethe-Heitler (BH) process at the Electron-Ion Collider (EIC) and the Electron-ion collider in China (EicC), using the azimuthal asymmetry of the final-state photons as an experimental probe of our theoretical predictions. Since the radiation occurs within the polarized background field, we incorporate the modified photon propagator to account for these effects. To the best of our knowledge, this study is the first to investigate bremsstrahlung radiation in polarized background fields within high-energy collider experiments, and we leverage azimuthal asymmetry to circumvent the experimental challenges associated with detecting polarized photons. Previous research has primarily focused on strong-field QED, electron radiation in intense laser fields \cite{DiPiazza2012,Mourou2006,Narozhny03072015}, and high-energy electron radiation in ordered crystals \cite{Baier_2005,Baier_2006,Uggerhoj2005,Azadegan2012}.

The remainder of this paper is organized as follows: Section \ref{2} presents the parameterization of photon propagators within polarized background fields, providing the explicit form of the propagators in such fields. Section \ref{3} offers numerical estimates of the azimuthal asymmetry in the Bethe-Heitler process at EIC and EicC, highlighting the impact of Coulomb corrections and Sudakov factors. Finally, Section \ref{4} concludes with a discussion of the implications of our findings and potential directions for future research.

\section{Photon propagators in polarized background fields}
\label{2}

In this study, we primarily investigate the BH process in the EIC and EicC. During the electron-ion collision, the electron absorbs a photon from the nuclear field and emits a final-state photon. As the electron traverses the external electromagnetic field of the nucleus, it undergoes multiple scatterings with the background field. Simultaneously, if the electron radiates a real photon, this process is also influenced by the background field. As mentioned earlier, the nuclear external photon field is polarized \cite{Li1,crosssection2,lg,lg2}, and the deceleration of the electron in this polarized background field, leading to photon emission, is the central theme of our study—bremsstrahlung in a polarized background field.

For the BH process in electron-ion collisions, it can be factorized into two parts: the hard part and the photon transverse momentum-dependent (TMD) distribution function describing the nuclear external photon distribution. In general, the photon correlator can be parameterized as follows \cite{lg,PhysRevD.83.105005,PhysRevLett.105.062001},
\begin{align}
 \left.\int\frac{2dy^-d^2y_\bot}{xP^+(2\pi)^3}e^{ik\cdot y}\langle P|F_{+\bot}^\mu(0)F_{+\bot}^\nu(y)|P\rangle\right|_{y^+=0}= \delta_\bot^{\mu\nu}f_1^\gamma\left(x,k_\bot^2\right) + \left(\frac{2k_\bot^\mu k_\bot^\nu}{k_\bot^2}-\delta_\bot^{\mu\nu}\right)h_1^{\bot\gamma}\left(x,k_\bot^2\right) 
\end{align}
where $\delta_\bot^{\mu\nu}=-g^{\mu\nu}+p^\mu n^\nu+p^\nu n^\mu$, and thus $k_\bot^2 = \delta_\bot^{\mu\nu}k_{\bot\mu}k_{\bot\nu}$. Here, $x$ represents the longitudinal momentum fraction carried by the photon. The functions $f_1^\gamma(x,k_\bot^2)$ and $h_1^{\bot\gamma}(x,k_\bot^2)$ are the unpolarized and linearly polarized photon TMDs, respectively.

Assuming that the nucleus moves along the $P^{+}$ direction, the dominant component of the gauge potential is $A^{+}$, while other components are suppressed by the Lorentz contraction factor $\gamma$. Based on this observation, after partial integration, the photon field strength tensor can be approximated as $F_{+\perp}^\mu F_{+\perp}^\nu \propto k_{\perp}^\mu k_{\perp}^\nu A^{+} A^{+}$, which leads to the relation \cite{Li1,Li2,eq,eq2},
\begin{equation}
    f_1^\gamma\left(x, k_{\perp}^2\right) = h_1^{\perp \gamma}\left(x, k_{\perp}^2\right)\label{e2}
\end{equation}
within the EPA, one obtains,
\begin{equation}
    x f_1^\gamma\left(x, k_{\perp}^2\right) = x h_1^{\perp \gamma}\left(x, k_{\perp}^2\right) = \frac{Z^2 \alpha_e}{\pi^2} k_{\perp}^2 \left[\frac{F\left(k_{\perp}^2 + x^2 M_p^2\right)}{\left(k_{\perp}^2 + x^2 M_p^2\right)}\right]^2\label{tmd}
\end{equation}
where $Z$ is the nuclear charge number, $F$ is the nuclear charge form factor, and $M_p$ is the proton mass. The nuclear charge form factor can be obtained from the STARlight Monte Carlo generator \cite{starlight},
\begin{equation}
    F(|k|) = \frac{4\pi\rho^0}{|k|^3A}\left[\sin\left(|k|R_A\right) - |k|R_A\cos\left(|k|R_A\right)\right]\frac{1}{a^2|k|^2+1}
\end{equation}
where $R_A=1.1A^{1/3} \text{ fm}$ and $a=0.7 \text{ fm}$. This parameterization closely resembles the Woods-Saxon distribution in phase space.

As the electron passes through the photon field, it undergoes multiple scatterings with the background photon field, leading to the Coulomb correction. This correction occurs either before or after photon radiation and can be absorbed into the photon correlator via a closed-loop gauge link. The explicit form is given by \cite{cc1,cc2,cc3},
\begin{equation}
\begin{aligned}
    \int\frac{dy^-d^2y_\bot}{P^+(2\pi)^3} & e^{ik\cdot y} 
      \left. \langle A | F_{+\bot}^\mu(0) U^\dag(0_\bot) U(y_\bot) F_{+\bot}^\nu(y) | A \rangle \right|_{y^+=0} 
     \\&= \int \frac{d^2y_\bot \, d^2y_\bot^\prime}{4\pi^3} e^{ik_\bot \cdot (y_\bot - y_\bot^\prime)} 
     \mathcal{F}^\mu(x, y_\bot) \mathcal{F}^{*\nu}(x, y_\bot^\prime) e^{i[\mathcal{V}(y_\bot) - \mathcal{V}(y_\bot^\prime)]}
     \\&=\frac{\delta_\bot^{\mu\nu}}{2}xf_1^{\prime\gamma}\left(x,k_\bot^2\right)+\left(\frac{k_\bot^\mu k_\bot^\nu}{k_\bot^2}-\frac{\delta_\bot^{\mu\nu}}{2}\right)xh_1^{\prime\bot\gamma}\left(x,k_\bot^2\right)
\end{aligned}
\end{equation}
where $U^\dag(0_\bot)U(y_\bot)$ represents the transverse gauge link, effectively incorporating the Coulomb correction from the hard part into the photon TMDs $f_1^{\prime\gamma}\left(x,k_\bot^2\right)$ and $xh_1^{\prime\bot\gamma}\left(x,k_\bot^2\right)$ . The explicit expression for the gauge link is.
\begin{equation}
    U(y_\bot) = \mathcal{P} e^{ie\int_{-\infty}^{+\infty} dz^- A^+(z^-,y_\bot)}
\end{equation}
Since photons are neutral, the correlator does not require a gauge link for gauge invariance. The potential can be further expressed as,
\begin{align}
    \mathcal{V}(y_\bot) \equiv e\int_{-\infty}^{+\infty} dz^- A^+(z^-,y_\bot)= \frac{\alpha Z}{\pi} \int d^2q_\bot e^{-iy_\bot\cdot q_\bot} \frac{F(q_\bot^2)}{q_\bot^2 + \delta^2}
\end{align}
where $\delta$ is the photon mass, used to regulate the infrared divergence, and $F$ is the nuclear charge form factor. The explicit form of the field strength in the correlator is.
\begin{align}
    \mathcal{F}^\mu(x,y_\bot) &\equiv \int_{-\infty}^{+\infty} dy^- e^{ixP^+y^-}F_{+\bot}^\mu(y^-,y_\bot) \\&= \frac{Ze}{4\pi^2} \int d^2q_\bot e^{-iy_\bot\cdot q_\bot} (iq_\bot^\mu) \frac{F(q_\bot^2+x^2M_p^2)}{q_\bot^2 + x^2M_p^2}\nonumber
\end{align}
The photon TMDs with final-state Coulomb corrections that we need can be obtained through orthogonality. We have.
\begin{align}
xf_1^{\prime\gamma}\left(x,k_\bot^2\right)=\int\ \frac{d^2y_\bot d^2y_\bot^\prime}{4\pi^3}e^{ik_\bot\cdot\left(y_\bot-y_\bot^\prime\right)}\mathcal{F}^\mu\left(x,y_\bot\right)\mathcal{F}^{\ast\nu}\left(x,y_\bot^\prime\right)\delta_\bot^{\mu\nu}e^{i\left[\mathcal{V}\left(y_\bot\right)-\mathcal{V}\left(y_\bot^\prime\right)\right]}
\end{align}
\begin{align}
xh_1^{\prime\bot\gamma}\left(x,k_\bot^2\right)=\int\ \frac{d^2y_\bot d^2y_\bot^\prime}{4\pi^3}e^{ik_\bot\cdot\left(y_\bot-y_\bot^\prime\right)}\mathcal{F}^\mu\left(x,y_\bot\right)\mathcal{F}^{\ast\nu}\left(x,y_\bot^\prime\right)\left(\frac{2k_\bot^\mu k_\bot^\nu}{k_\bot^2}-\delta_\bot^{\mu\nu}\right)e^{i\left[\mathcal{V}\left(y_\bot\right)-\mathcal{V}\left(y_\bot^\prime\right)\right]}
 \end{align}
As shown in Eq. \ref{e2}, we find that the photons surrounding the atomic nucleus are linearly polarized. Here, not only does $f_1^\gamma(x, k_\bot^2)$ exist, but also $h_1^{\bot\gamma}(x, k_\bot^2)$, which represents the distribution function of linearly polarized photons. When it acts as a background field, the photon propagator becomes more complicated, taking the following form (see Appendix \ref{App1} for more details),
\begin{equation}
\begin{aligned}
    &\left\langle B\middle|\ A_\mu\left(x\right)A_\nu\left(y\right)\middle|\ B\right\rangle
\\&=\int{\frac{d^4 k}{(2\pi)^4}e^{-ik\cdot\left(x-y\right)}\frac{-i}{k^2+i\varepsilon}}2E_k (2\pi)^3\frac{k_z}{E_k}[g_{\bot\mu\nu}{f}_1^\gamma\left(E_k,k_\bot^2\right)-(g_{\bot\mu\nu}+\frac{2k_{\bot\mu}k_{\bot\nu}}{k_\bot^2}){h}_1^{\bot\gamma}(E_k,k_\bot^2)]  \label{pro}
\end{aligned}
 \end{equation}
where $B$ denotes the polarized photon background field, and $-g_\bot^{\mu\nu} = \delta_\bot^{\mu\nu} = -g^{\mu\nu} + p^\mu n^\nu + p^\nu n^\mu$. As shown in Ref.~\cite{Li1} and Eq.~(\ref{tmd}), the polarized and unpolarized photon distribution functions
are identical. Their energy-dependent expressions within the textbook-level EPA \cite{eq} are given by,
\begin{equation}
\begin{aligned}
f_1^\gamma(E_k, k_{\perp}^2) = h_1^{\perp \gamma}(E_k, k_{\perp}^2) = \frac{Z^2 \alpha_e}{\pi^2} \frac{k_\perp^2}{E_k}\left[\frac{F(k_\perp^2 + E_k^2/\gamma^2)}{(k_\perp^2 + E_k^2/\gamma^2)}\right]^2
\end{aligned}
\end{equation}
where $\gamma$ is the Lorentz contraction factor. The final-photon energy and longitudinal momentum are related to the transverse momentum and rapidity through \( E_k = k_\perp \cosh y_\gamma \) and \( k_z = k_\perp \sinh y_\gamma \). When studying the motion of photons in the background field, the photon propagator in the polarized background field depends on both the polarized and unpolarized photon distribution functions in the field. This equation provides a reformulated description, within the framework of the parton model in field theory, of how the background field affects bremsstrahlung photons within it. Furthermore, it shows that in a linearly polarized background field, the bremsstrahlung photon beam is linearly polarized, and its spectrum corresponds to the photon distribution function in the background field.

\section{Observables}
\label{3}
We can investigate bremsstrahlung radiation in a polarized background field through the BH process at EIC and EicC.
\begin{equation}
    e(k_1) + \gamma(k_2) \longrightarrow e(p_1) + \gamma(p_2)
\end{equation}
In this process, an electron undergoes scattering with a heavy nucleus, absorbing a photon from the nuclear Coulomb field and subsequently emitting a real photon. Since the emitted photon originates from interactions with the external Coulomb field of the nucleus, the corresponding final-state photon propagator is modified according to the Eq. (\ref{pro}). The azimuthal asymmetry of the final-state particles can serve as an experimental signature to verify the polarization characteristics of the emitted photon beam.

The resummation of all large logarithmic terms induced by final-state soft photon radiation results in the Sudakov factor, given by    $e^{-\frac{\alpha_e}{2\pi} \ln^2\left(\frac{P_\perp^2}{\mu_r^2}\right)}$,where $\mu_r = 2 e^{-\gamma_E}/|r_\perp|$ \cite{cc3}. After convoluting the Sudakov factor over the impact parameter space, we obtain,
\begin{equation}
\begin{aligned}
    &\frac{d\sigma}{dy_\gamma d^2P_\bot d^2q_\bot}=\int\ \frac{d^2r_\bot}{(2\pi)^2}e^{ir_\bot\cdot q_\bot}e^{-\frac{\alpha_e}{2\pi}\ln^2\frac{Q^2}{\mu_r^2}}\int\ d^2k_\bot e^{ir_\bot\cdot k_\bot}2E_k (2\pi)^3\frac{k_z}{E_k}[H_1xf_1^\gamma\left(x,k_{2\bot}^2\right){f}_1^\gamma\left(E_k,p_{2\bot}^2\right)\\&
+H_2{xh}_1^{\bot\gamma}\left(x,k_{2\bot}^2\right){f}_1^\gamma\left(E_k,p_{2\bot}^2\right)+H_3{xf}_1^\gamma\left(x,k_{2\bot}^2\right)h_1^{\bot\gamma}\left(E_k,p_{2\bot}^2\right)+H_4{xh}_1^{\bot\gamma}(x,k_{2\bot}^2)h_1^{\bot\gamma}(E_k,p_{2\bot}^2)]
\end{aligned}
\end{equation}
where,
\begin{equation}
\begin{aligned}
H_1&=\frac{2z^2\left(3z^2-4z+2\right)}{p_\bot^4};\\
H_2&=\frac{4\left(z-1\right)z^3cos\left(2\phi\right)}{p_\bot^4};\\
H_3&=\frac{4\left(z-1\right)z^3}{p_\bot^4};\\
H_4&=\frac{4\left(z-1\right)^2z^2cos{\left(2\phi\right)}}{p_\bot^4}
\end{aligned}
\end{equation}
where $z$ denotes the longitudinal momentum fraction of the incoming electron carried by the final-state photon. Since the total transverse momentum $q_{\perp}$ satisfies $q_{\perp} = k_{2\perp} = p_{1\perp} + p_{2\perp} \approx 0$, we approximate $p_{1\perp} \approx -p_{2\perp} \approx P_\perp$. The azimuthal angle $\phi$ is defined as the angle between $P_\perp$ and the total transverse momentum $k_{2\perp}$.

\begin{figure*}[htbp]
    \centering
    \begin{subfigure}{0.5\textwidth}
        \centering
        \includegraphics[width=\textwidth]{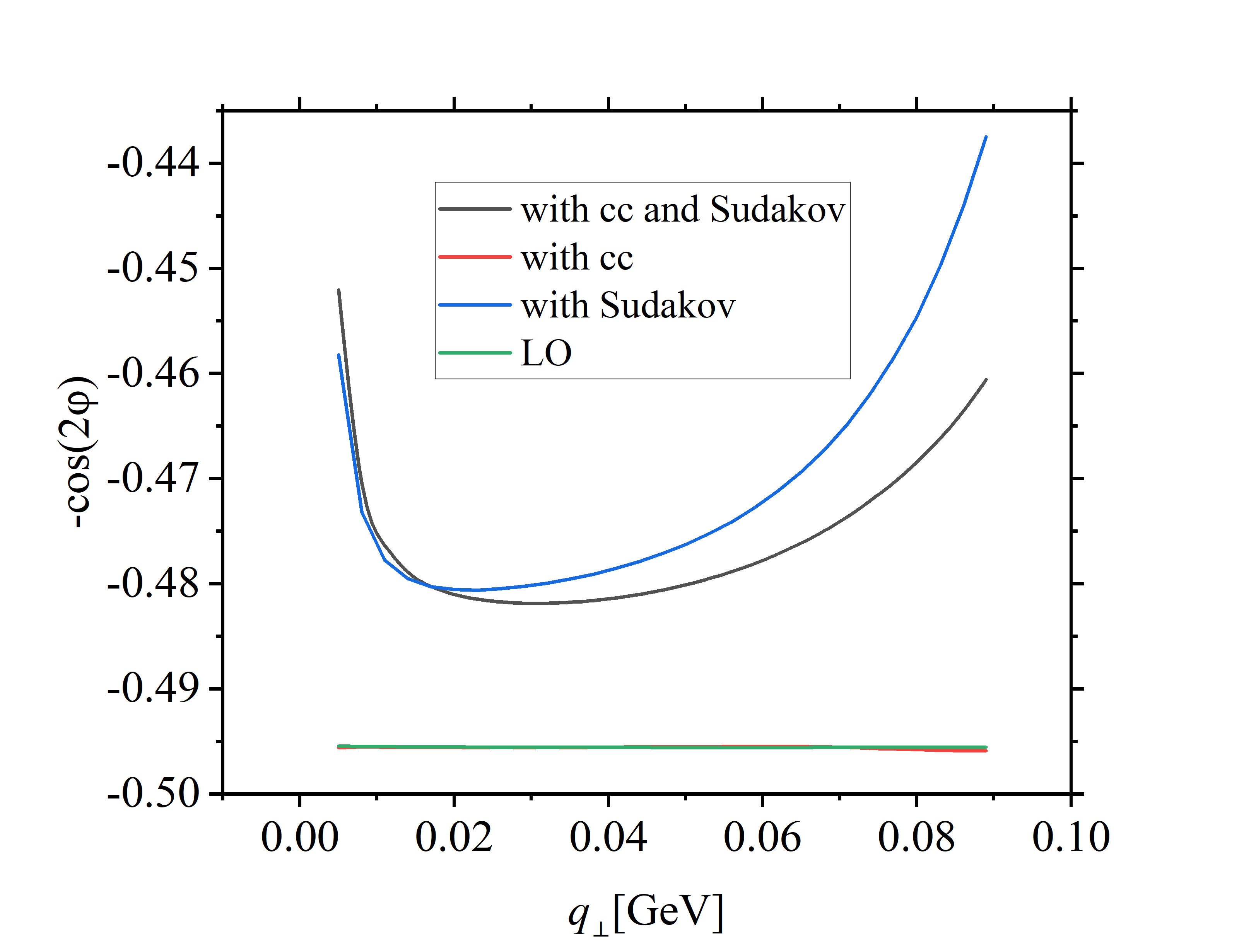} 
    \end{subfigure}\hfill
    \begin{subfigure}{0.5\textwidth}
        \centering
        \includegraphics[width=\textwidth]{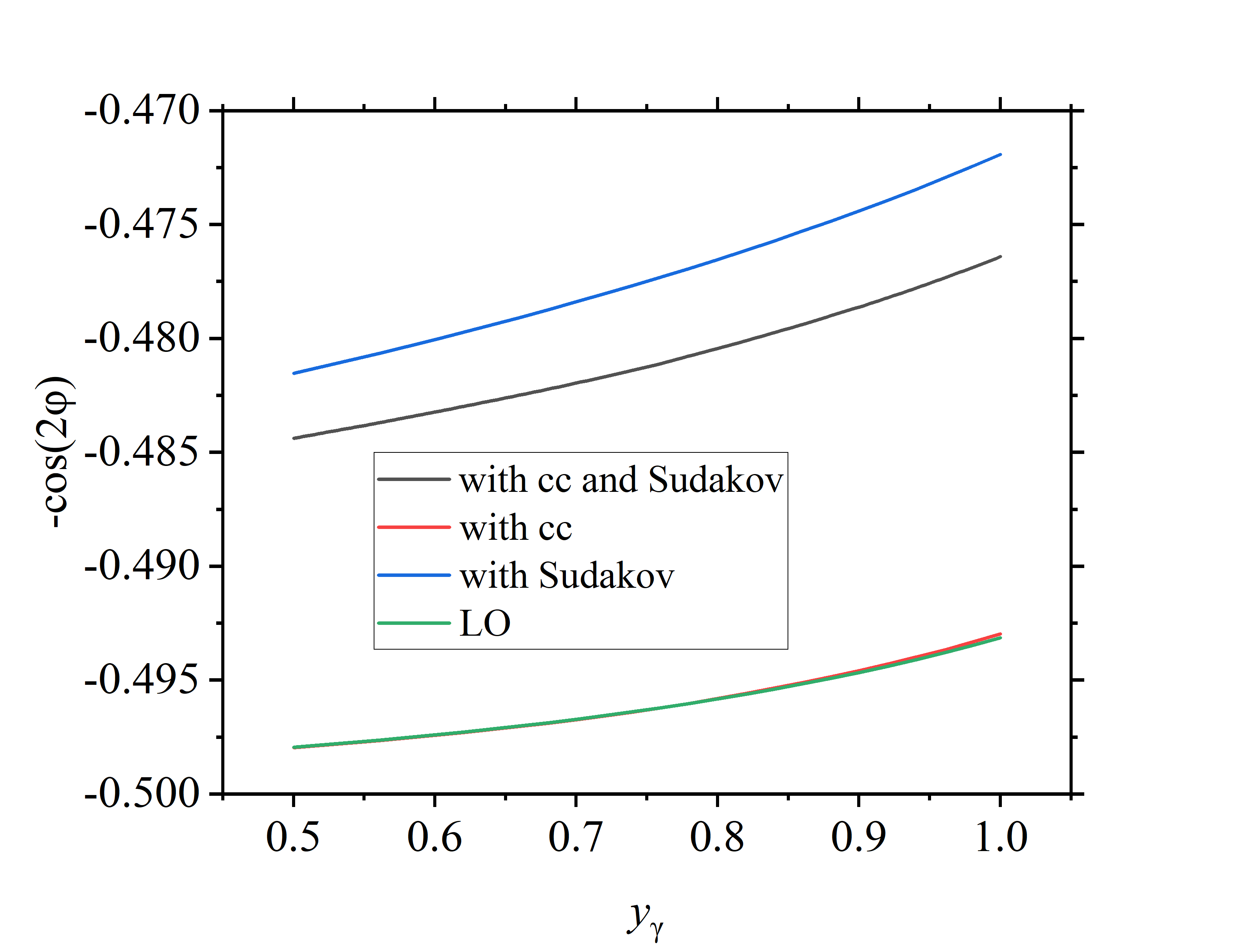} 
    \end{subfigure}
    \caption{The azimuthal asymmetry as a function of $q_{\perp}$ (left panel) and $y_\gamma$ (right panel) with and without taking into account the Coulomb corrections (cc) or Sudakov for a Pb target at EicC. The asymmetry is averaged over the $P_{\perp}$ region $[0.3~\mathrm{GeV}, 0.4 ~\mathrm{GeV}]$. In the left plot, the emitted photon rapidity $y_\gamma$ is integrated over the region $[0.5,1]$. In the right plot, the total transverse momentum $q_{\perp}$ is fixed to be 50 MeV .}
    \label{f.fig1}
\end{figure*}

\begin{figure*}[htbp]
    \centering
    \begin{subfigure}{0.5\textwidth}
        \centering
        \includegraphics[width=\textwidth]{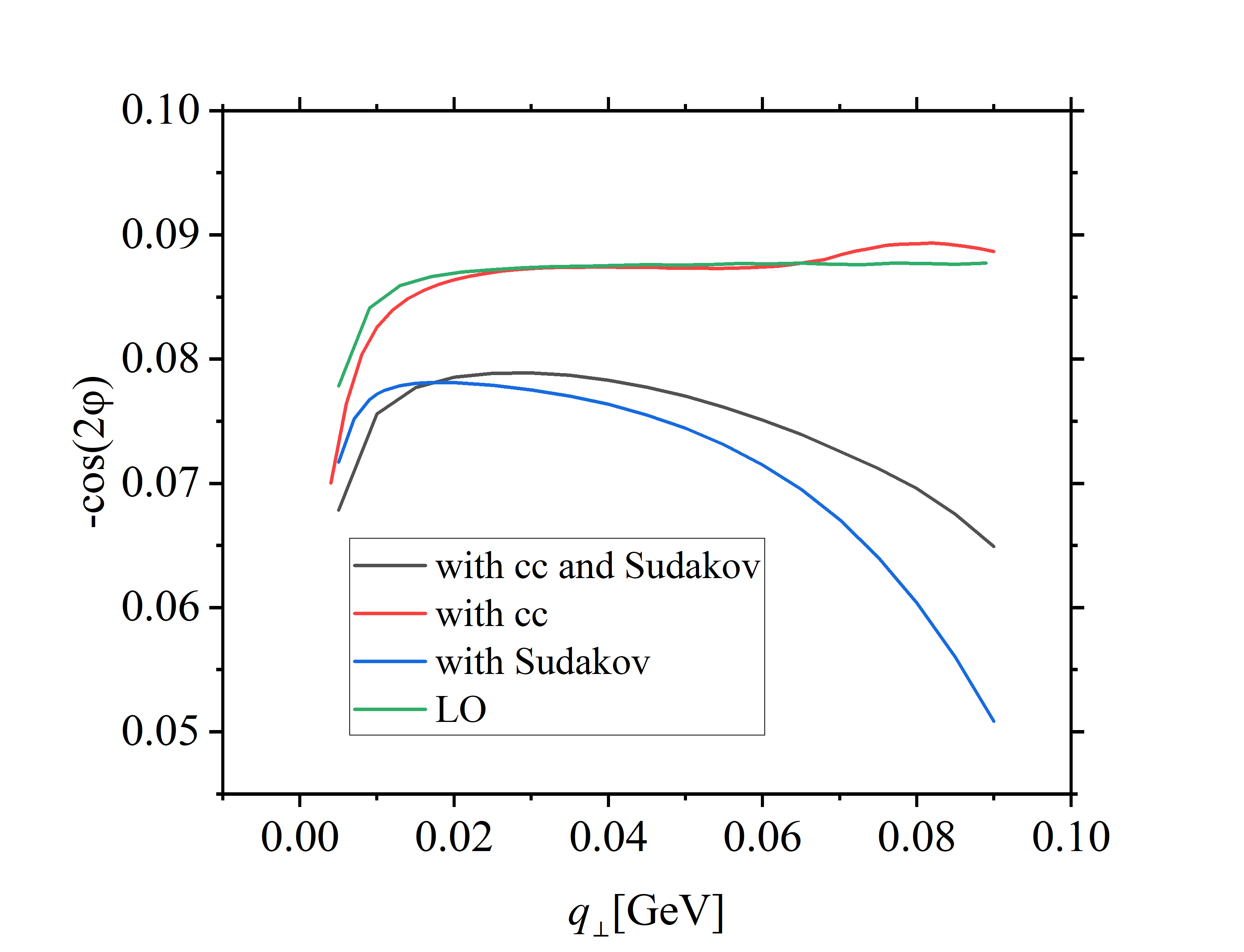} 
    \end{subfigure}\hfill
    \begin{subfigure}{0.5\textwidth}
        \centering
        \includegraphics[width=\textwidth]{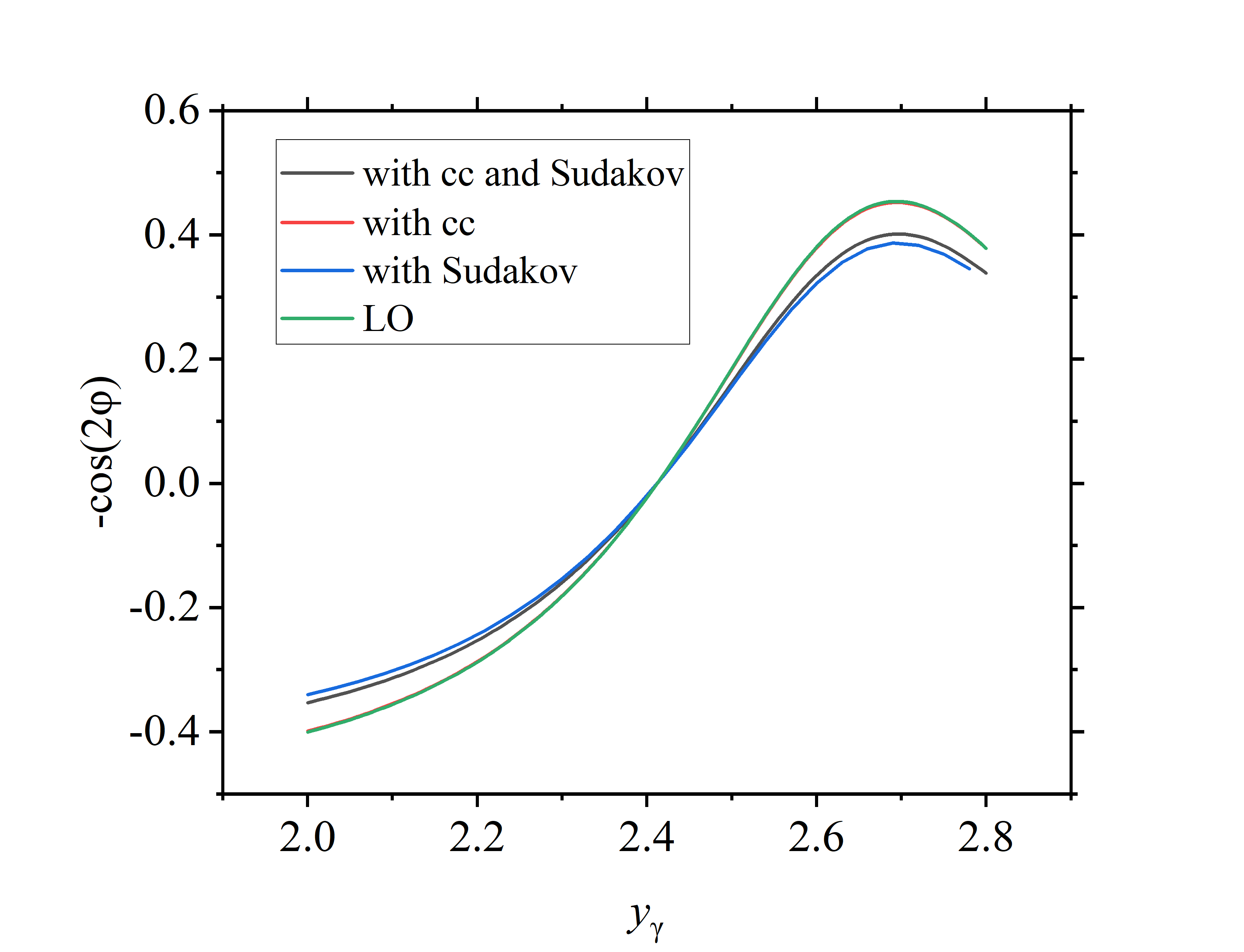} 
    \end{subfigure}
    \caption{The azimuthal asymmetry as a function of $q_{\perp}$ (left panel) and $y_\gamma$ (right panel) with and without taking into account the Coulomb corrections (cc) or Sudakov for a Pb target at EIC. The asymmetry is averaged over the $P_{\perp}$ region $[1.5 ~\mathrm{GeV}, 2 ~\mathrm{GeV}]$. In the left plot, the emitted photon rapidity $y_\gamma$ is integrated over the region $[2,2.8]$. In the right plot, the total transverse momentum $q_{\perp}$ is fixed to be 50 MeV .}
    \label{f.fig2}
\end{figure*}

Now, we proceed to numerically evaluate the magnitude of this effect at both EIC and EicC. Specifically, we analyze the dependence of the $\cos 2\phi$ azimuthal asymmetry on the rapidity of the final-state photon as well as on the total transverse momentum at the energy scales of EIC and EicC. The beam energies in these two colliders are as follows: for EIC, the electron and heavy-ion beam energies are 18 GeV and 100 GeV, respectively, while for EicC, they are 3.5 GeV and 8 GeV, respectively. The azimuthal asymmetry, quantified by the expectation value $\langle \cos{2\phi} \rangle$, is given by,
\begin{align}
    \left\langle\cos{\left(2\phi\right)}\right\rangle=\frac{\int\ \frac{d\sigma}{d\mathcal{P}.\mathcal{S}.}\cos{2\phi d\mathcal{P}}.\mathcal{S}.}{\int\ \frac{d\sigma}{d\mathcal{P}.\mathcal{S}.}d\mathcal{P}.\mathcal{S}.}
\end{align}
where $d\mathcal{P}.\mathcal{S}.$ is the phase space factor. In the context of EicC and EIC, we calculate the azimuthal asymmetry $-\cos(2\phi)$ as a function of transverse momentum $q_{\perp}$ and photon rapidity $y_{\gamma}$ in the bremsstrahlung process. We also investigate the effects of Coulomb corrections and Sudakov suppression. The numerical results are presented in Fig.~\ref{f.fig1} and Fig.~\ref{f.fig2}.  

In Fig.~\ref{f.fig1}, we consider the kinematic settings for EicC, where the transverse momentum of the parent particle is chosen within the range $P_{\perp} \in [0.3, 0.4]$ GeV. In the left panel, we observe that $-\cos(2\phi)$ exhibits a clear suppression effect due to Sudakov corrections: the result including Sudakov effects (blue line) is significantly reduced in the low-$q_{\perp}$ region compared to the cases without Sudakov corrections (red and black lines). In the right panel, as $y_{\gamma}$ increases, the azimuthal asymmetry shows a slight enhancement. The Coulomb correction is negligible before the Sudakov effect is taken into account (green and red lines). After the Sudakov effect is taken into account, its effect becomes slightly larger (blue and black lines).  

In Fig.~\ref{f.fig2}, we switch to the EIC kinematics, choosing a higher transverse momentum range for the parent particle, $P_{\perp} \in [1.5,2]$ GeV. In the left panel, $-\cos(2\phi)$ is relatively small at low $q_{\perp}$ but increases rapidly with $q_{\perp}$. The Sudakov effect is particularly pronounced at large $q_{\perp}$. Similarly, in the right panel, $-\cos(2\phi)$ increases with $y_{\gamma}$, reaching a peak around $y_{\gamma} \approx 2.65$ before slightly decreasing. This suggests a non-monotonic dependence of azimuthal asymmetry on photon rapidity. The Coulomb correction is negligible before the Sudakov effect is considered. After the Sudakov effect is considered, its effect is slightly larger. The same findings are found here as in the EicC case. The Coulomb correction is negligible before the Sudakov effect is taken into account (green and red lines, the red is overlaid by green). After the Sudakov effect is taken into account, its effect becomes slightly larger (blue and black lines). These findings provide valuable insights for future experimental measurements of azimuthal asymmetry at EIC and EicC.

\section{Summary and Outlook}
\label{4}
In summary, we have developed a consistent quantum-field-theoretical framework to describe bremsstrahlung radiation from an electron propagating in a polarized background photon field. 
Starting from the photon two-point correlator, we constructed the background-modified photon propagator in terms of TMD distributions. 
Coulomb effects were incorporated through a gauge-link formalism, while soft-photon emissions were resummed via the Sudakov factor. 
This formulation provides an analytic expression for the polarized BH spectrum, linking the photon dynamics in the background field directly to measurable observables.

Our numerical calculations under EIC and EicC kinematics demonstrate that the emitted photon beam inherits the linear polarization of the background field, leading to a $\cos 2\phi$-type azimuthal asymmetry. 
The Sudakov suppression dominates at low transverse momentum, whereas Coulomb corrections induce a mild enhancement in the high-$q_\perp$ region. 
The predicted asymmetry, with a magnitude of several percent, should be experimentally accessible in forthcoming measurements of ultra-peripheral electron--ion collisions.

These results indicate that the polarization-dependent modulation provides sensitivity to the underlying photon polarization in the nuclear field.
The framework established here may also be extended to other QED processes in polarized media or to nuclear environments where electromagnetic field polarization plays a dynamical role.

\appendix

\section{Derivation of the Photon Propagator in a Polarized Background Field}
\label{App1}

The standard photon propagator in vacuum is defined as $\langle 0 | A_\mu(x) A_\nu(y) | 0 \rangle$. When the vacuum is replaced by a photon background field, the corresponding photon propagator is modified to $\langle B | A_\mu(x) A_\nu(y) | B \rangle$. In particular, when the background field is provided by the photon flux in ultraperipheral heavy-ion collisions, the photon propagator can be parameterized as,
\begin{equation}
    \begin{aligned}
&\left\langle B \middle| A_\mu(x)\, A_\nu(y) \middle| B \right\rangle \\
&= \int \frac{d^4k}{(2\pi)^4} \, e^{-ik\cdot(x - y)} \, \frac{-i}{k^2 + i\varepsilon} \, 2E_k(2\pi)^3 \frac{k_z}{E_k}\left[f_1^\gamma(E_k, k_\perp^2) g_{\perp}^{\mu\nu}- h_1^{\perp\gamma}(E_k, k_\perp^2)  \left( g_{\perp}^{\mu\nu} + \frac{2 k_\perp^\mu k_\perp^\nu}{k_\perp^2} \right)\right]
\label{p4}
\end{aligned}
\end{equation}
where the $E_k$ and $k_z$ are the photon's energy and longitudinal momentum. To derive this expression, we present a detailed calculation of the photon two-point function in a polarized background field—representative of the electromagnetic (EM) field in peripheral heavy-ion collisions—using a mixed-state formalism based on a polarization density matrix.

\subsection*{1 Photon Two-Point Function and Density Matrix}

The background-field-modified photon two-point function is defined as,
\begin{equation}
\left\langle B \middle| A_\mu(x)\, A_\nu(y) \middle| B \right\rangle = \mathrm{Tr} \left[ \rho\, A_\mu(x)\, A_\nu(y) \right]
\end{equation}
where \( \rho \) is the photon polarization density matrix. For an ensemble of photons with definite momentum and polarization, this density matrix takes the form,
\begin{equation}
\rho = \sum_{\lambda,\lambda'} \int \frac{d^3k}{(2\pi)^3 2E_k}\, \eta_{\lambda\lambda'}(k)\, |k,\lambda\rangle \langle k,\lambda'|
\end{equation}
where \( \eta_{\lambda\lambda'}(k) \) encodes the occupation and polarization structure. Inserting this into the correlator yields,
\begin{align}
\left\langle B \middle| A_\mu(x)\, A_\nu(y) \middle| B \right\rangle
= \sum_{\lambda,\lambda'} \int \frac{d^3k}{(2\pi)^3 2E_k}\, \eta_{\lambda\lambda'}(k)\, \epsilon_\lambda^\mu(k)\, \epsilon_{\lambda'}^{*\nu}(k)\, e^{-ik \cdot (x - y)}
\end{align}
where \( \epsilon_\lambda^\mu(k) \) are the polarization vectors, and $ \left. A_\nu(y) \middle| k, \lambda \right\rangle=\left.\epsilon_{\lambda\nu}(k) e^{-ik \cdot y}\middle| 0 \right\rangle$.

\subsection*{2 Structure of the Polarization Density Matrix}

In the transverse polarization basis \( \lambda = 1,2 \), corresponding to \( \epsilon_1 = \epsilon_x \), \( \epsilon_2 = \epsilon_y \), the polarization density matrix can be expanded in terms of Pauli matrices.
\begin{equation}
\eta_{\lambda\lambda'}(k) =  f(k)\, \delta_{\lambda\lambda'} +  \sum_{i=1}^{3} h^i(k)\, \sigma^i_{\lambda\lambda'}
\end{equation}
Here, \( f(k) \) describes the unpolarized photon distribution, and \( h^i(k) \) are polarization parameters analogous to the Stokes parameters. \( h^3(k) \) encodes the difference between \( \epsilon_x \) and \( \epsilon_y \) intensity (linear polarization along \( x \) vs. \( y \)). \( h^1(k) \) characterizes coherence between \( \epsilon_x \) and \( \epsilon_y \) (i.e., linear polarization along \( x \pm y \)). \( h^2(k) \) represents circular polarization (phase difference between \( \epsilon_x \) and \( \epsilon_y \)). The $h^2$ is neglected here, as the background field is dominantly linearly polarized in ultraperipheral heavy-ion collisions \cite{Li1,Li2}.

Contracting \( \eta_{\lambda\lambda'} \) with polarization vectors, we obtain.
\begin{align}
\sum_{\lambda,\lambda'} \eta_{\lambda\lambda'}(k)\, \epsilon_\lambda^\mu(k)\, \epsilon_{\lambda'}^{*\nu}(k)
=  f(k) \sum_\lambda \epsilon_\lambda^\mu(k)\, \epsilon_\lambda^{*\nu}(k)
+  \sum_{i=1}^{3} h^i(k) \sum_{\lambda,\lambda'} \sigma^i_{\lambda\lambda'}\, \epsilon_\lambda^\mu(k)\, \epsilon_{\lambda'}^{*\nu}(k)
\end{align}
The polarization sum gives $\sum_{\lambda=1,2} \epsilon_\lambda^\mu(k)\, \epsilon_\lambda^{*\nu}(k) = -g_\perp^{\mu\nu}$. Evaluating the Pauli matrix term \( \sigma^3 \) gives,
\begin{align}
\sum_{\lambda,\lambda'} \sigma^3_{\lambda\lambda'}\, \epsilon_\lambda^\mu\, \epsilon_{\lambda'}^{*\nu}& = \epsilon_x^\mu \epsilon_x^{*\nu} - \epsilon_y^\mu \epsilon_y^{*\nu}=2\epsilon_x^\mu \epsilon_x^{*\nu}- (\epsilon_x^\mu \epsilon_x^{*\nu}+\epsilon_y^\mu \epsilon_y^{*\nu})=  \frac{2k_\perp^\mu k_\perp^\nu}{k_\perp^2}+g_\perp^{\mu\nu}
\end{align}
 where the linear polarization vectors can be chosen such that \( \epsilon_x^\mu(k) \) is aligned with the transverse momentum vector \( k_\perp^\mu \), i.e., $\epsilon_x^\mu(k)=k_\perp^\mu/|k_\perp|$. Hence, the tensor structure associated with \( h^3(k) \sigma^3 \) corresponds to a net linear polarization aligned with the $k_\perp$. The \( \sigma^1 \) term gives,
\begin{align}
\sum_{\lambda,\lambda'} \sigma^1_{\lambda\lambda'}\, \epsilon_\lambda^\mu\, \epsilon_{\lambda'}^{*\nu} = \epsilon_x^\mu \epsilon_y^{*\nu} + \epsilon_y^\mu \epsilon_x^{*\nu}
\end{align}
which describes interference between orthogonal linear polarization directions. The \( \sigma^2 \) term, corresponding to circular polarization, is neglected within the EPA. Therefore, the full contraction becomes.
\begin{align}
\sum_{\lambda,\lambda'} \eta_{\lambda\lambda'}(k)\, \epsilon_\lambda^\mu(k)\, \epsilon_{\lambda'}^{*\nu}(k)&= f(k)\, (-g_\perp^{\mu\nu})+  h^3(k)\left( g_\perp^{\mu\nu} + \frac{2 k_\perp^\mu k_\perp^\nu}{k_\perp^2} \right)+  h^1(k)\left( \epsilon_x^\mu \epsilon_y^{*\nu} + \epsilon_y^\mu \epsilon_x^{*\nu} \right)\\
&= f(k)\, (-g_\perp^{\mu\nu})+  h^3(k)\left( g_\perp^{\mu\nu} + \frac{2 k_\perp^\mu k_\perp^\nu}{k_\perp^2} \right)\notag
\end{align}
Within the EPA framework, photons are assumed to be linearly polarized along a fixed transverse direction—typically radial with respect to the emitting nucleus—and the polarization state is pure for each \( k_\perp \). As a result, coherence terms between orthogonal directions vanish, implying \( h^1(k) = 0 \), which is also evident from Refs. \cite{Li1} and \cite{Li2}.
 
\subsection*{3 The Final Form}

The final form of the photon two-point function in a polarized background field is given by.
\begin{equation}
    \begin{aligned}
&\left\langle B \middle| A_\mu(x)\, A_\nu(y) \middle| B \right\rangle \\
&= \int \frac{d^3k}{(2\pi)^3 2E_k} \, e^{-ik\cdot(x - y)}\left[ f(k)\, (-g_\perp^{\mu\nu}) + h^3(k)\left( g_\perp^{\mu\nu} + \frac{2 k_\perp^\mu k_\perp^\nu}{k_\perp^2} \right) \right] \\
&= \int \frac{d^3k}{(2\pi)^3 2E_k} \, e^{-ik\cdot(x - y)} 
\left[-  f_1^\gamma(k_z, k_\perp^2)  2E_k(2\pi)^3\, g_{\perp}^{\mu\nu}+  h_1^{\perp\gamma}(k_z, k_\perp^2)  2E_k(2\pi)^3 \left( g_{\perp}^{\mu\nu} + \frac{2 k_\perp^\mu k_\perp^\nu}{k_\perp^2} \right) \right] \\
&= \int \frac{d^4k}{(2\pi)^4} \, e^{-ik\cdot(x - y)} \, \frac{-i}{k^2 + i\varepsilon} 
\left[ f_1^\gamma(k_z, k_\perp^2)  2E_k(2\pi)^3\, g_{\perp}^{\mu\nu}
- h_1^{\perp\gamma}(k_z, k_\perp^2)  2E_k(2\pi)^3 \left( g_{\perp}^{\mu\nu} + \frac{2 k_\perp^\mu k_\perp^\nu}{k_\perp^2} \right) \right]
\label{p1}
\end{aligned}
\end{equation}
In the third line of the above equation, the momentum-space photon distribution function \( f(k) \) and the three-momentum distribution \( f_1^\gamma(k_z, k_\perp) \) differ by a normalization factor of \( 2E_k (2\pi)^3 \), which arises from the relativistic normalization of one-photon states, $\langle k | k' \rangle = (2\pi)^3 2E_k \, \delta^{(3)}(\mathbf{k} - \mathbf{k}')$. Furthermore, the \( f_1^\gamma(E_k, k_\perp) \), which depends on the photon energy, is related to the three-momentum distribution 
\( f_1^\gamma(k_z, k_\perp) \) through

\begin{equation}
f_1^\gamma(k_z, k_\perp^2)= f_1^\gamma(E_k, k_\perp^2) \frac{k_z}{E_k}\label{p2}
\end{equation}
where the factor \( \frac{dE_k}{dk_z} = \frac{k_z}{E_k} \) represents the Jacobian arising from the transformation between energy and longitudinal-momentum variables,. A similar reasoning applies to the linearly polarized photon distribution, leading to
\begin{equation}
h^3(k) = h_1^{\perp \gamma}(k_z, k_\perp^2) 2E_k(2\pi)^3
\end{equation}
and
\begin{equation}
h_1^{\perp\gamma}(k_z, k_\perp^2) = h_1^{\perp \gamma}(E_k, k_\perp^2) \frac{k_z}{E_k}.\label{p3}
\end{equation}
This change of variables ensures that the photon propagator is expressed in terms of physically meaningful TMD distributions consistent with the EPA formalism. In the last line of Eq.~(\ref{p1}), the transition from the three-dimensional integral to the four-dimensional one uses the residue theorem,
\begin{align}
    \int \frac{d^4k}{(2\pi)^4} \frac{i}{k^2 + i\varepsilon}\cdots=\int \frac{d^3k}{(2\pi)^3 2E_k}\cdots
\end{align}
which effectively captures the contribution of on-shell photon propagation in the background field. The final result clearly separates the contributions from unpolarized photons and those with linear polarization aligned with \( k_\perp \), and omits coherence terms, consistent with the assumptions underlying the EPA. In summary, by combining Eqs.~(\ref{p1}), (\ref{p2}), and (\ref{p3}), the photon propagator in the polarized background field can be expressed as Eq.~(\ref{p4}).

\bibliographystyle{unsrt}
\bibliography{ref}

\end{document}